\documentclass[singlecolumn,secnumarabic,amssymb, nobibnotes, aps, prd,longbibliography]{revtex4-2}
\usepackage{amsmath,amssymb,amsfonts}
\usepackage{graphicx}
\usepackage{hyperref}
\usepackage{bm}
\usepackage{lineno}
\usepackage{color}

\newcommand{\env}[1]{\texttt{#1}}
\begin{document}
\title{Phase-Space Topology and Spectral Flow in Screened Magnetized Plasmas}
\author{Xianhao Rao}
\author{Adil Yolbarsop}\email{adil0608@ustc.edu.cn; yolbarsop@gmail.com}
\author{Hong Li}\email{honglee@ustc.edu.cn}
\author{Wandong Liu}
\affiliation{School of Nuclear Science and Technology, University of Science and Technology of China. No. 443, Huangshan Road, Hefei, Anhui, China.}

\begin{abstract}

Topological wave phenomena in continuous media are fundamentally challenged by unbounded spectra and the absence of a compact Brillouin zone, which obstruct conventional bulk--interface formulations. We develop a unified phase-space framework for screened magnetized plasma based on a pseudo-Hermitian formulation with a positive-definite metric, enabling a generalized Schrödinger description and a Weyl-symbol analysis of the bulk generator. We show that the bulk symbol hosts isolated band degeneracies acting as Berry--Chern monopoles, including a higher-order spin-1 degeneracy with topological charge $+2$ that generically splits into two spin-$\tfrac{1}{2}$ Weyl points under symmetry breaking. To characterize topology in this noncompact setting, we introduce a strip-gap Chern number associated with finite real-frequency strips of the bulk spectrum, extending band Chern topology to continuum systems. This invariant governs the spectral flow of interface modes induced by spatial variations of the magnetic field and establishes a bulk--interface correspondence at the level of phase-space symbols. By solving the interface eigenvalue problem, we demonstrate that the net spectral flow across the strip gap is determined by the enclosed monopole charge. We further show that this correspondence persists under collisional damping, provided that a finite strip gap remains and no exceptional points enter it. Our results provide a systematic phase-space framework for topological wave transport in continuous media beyond compact-band and idealized Hermitian settings.

\end{abstract}
\maketitle
\section{Introduction}
Topological phases of waves have emerged as a unifying concept across
condensed matter physics~\cite{hasan2010colloquium,qi2011topological},
photonics~\cite{lu2014topological,ozawa2019topological},
acoustics~\cite{huber2016topological,susstrunk2015observation},
fluid dynamics~\cite{delplace2017topological,graf2021topology}, and plasma physics~\cite{fu2021topological,parker2020topological,qin2023topological,rao2025symmetry}
revealing that unidirectional transport protected by band topology can arise from global features of the underlying spectrum rather than microscopic details.
While early studies focused primarily on lattice systems with discrete
Bloch bands~\cite{haldane1988model,thouless1982quantized}, it has become increasingly clear that similar topological phenomena can also occur in continuous media, including fluids, plasmas, and geophysical flows~\cite{delplace2017topological,graf2021topology,silveirinha2015chern,perez2025topology}.
Prominent examples include equatorial waves in rotating fluids, magnetized plasma waves, and electromagnetic continua, where topology manifests itself through spectral flow and protected interface modes rather than conventional edge states~\cite{parker2020topological,graf2013bulk,hatsugai1993chern}.

Despite this progress, establishing a systematic bulk--interface correspondence
for continuous media remains challenging~\cite{silveirinha2015chern,parker2020topological,graf2021topology}.
Unlike lattice systems, continuous media generally lack a compact Brillouin
zone and exhibit unbounded spectra, rendering standard band-topological
invariants inapplicable~\cite{silveirinha2015chern,graf2013bulk}.
In particular, the notion of a band gap must be generalized to accommodate spectral continua, and the definition of topological invariants requires careful treatment in phase space~\cite{silveirinha2015chern,kellendonk2002edge}.
These difficulties are further exacerbated in plasma systems, where the dynamical variables are subject to constraints such as Gauss's law and screening, leading to non-canonical Hamiltonian structures.~\cite{morrison1998hamiltonian}.

In this work, we address these challenges by formulating the dynamics of a
screened magnetized plasma as a pseudo-Hermitian system with a positive-definite
metric.
This structure allows the linearized equations to be cast into a generalized Schr\"odinger form with a real spectrum in the absence of dissipation~\cite{mostafazadeh2002pseudo}.
By employing the Weyl--Wigner formalism, we analyze the Weyl symbol of the effective generator and identify isolated band degeneracies in phase space~\cite{wigner1932quantum,littlejohn1986semiclassical}.
To characterize topology in this continuous setting, we introduce a strip-gap Chern number, defined for frequency intervals free of spectrum at large phase-space distances, which naturally generalizes the concept of a band gap in lattice systems~\cite{silveirinha2015chern,parker2020topological,gong2018topological}.

From a broader perspective, bulk--interface correspondence in continuous media can be understood directly at the level of phase-space symbols. For Hermitian systems with a spectral gap, the spectral flow of the corresponding operator is governed by the Chern number of the Weyl symbol~\cite{faure2019manifestation,jezequel2023mode}.
Recent developments have shown that, for spin-$\tfrac{1}{2}$ systems, this correspondence persists even in non-Hermitian settings, where Berry monopoles associated with complex-energy degeneracies control the spectral flow~\cite{jezequel2023non}. More fundamentally, the presence of a gap allows one to define a topological invariant through the spectral projector of the symbol bands, providing a concrete setting to explore bulk--interface correspondence in higher-spin systems, including pseudo-Hermitian and weakly non-Hermitian regimes in which a real strip gap remains well defined.

Within this context, the screened magnetized plasma considered here realizes a
pseudo-Hermitian continuum system that hosts a spin-1 band degeneracy at
$k_z = 0$.
This makes it an ideal platform to explore and validate a generalized bulk--interface correspondence beyond the conventional spin-$\tfrac{1}{2}$ paradigm.

We show that the bulk symbol of the screened plasma supports a higher-order
spin-1 degeneracy carrying a topological charge of $+2$, which splits into two
spin-$\tfrac{1}{2}$ Weyl points under symmetry breaking (\(k_z=0\longrightarrow k_z\neq0\)).
The associated gap Chern number governs the spectral flow of interface modes
when the background magnetic field varies spatially.
By solving the corresponding one-dimensional eigenvalue problem, we numerically confirm a bulk–interface correspondence in which the net number of chiral modes traversing the strip gap is dictated by the total monopole charge enclosed by the path of $\mu(x)$ in parameter space.
Our results provide a unified topological interpretation of interface waves in
screened plasmas and connect to known descriptions of equatorial geostrophic waves in the long-wavelength limit~\cite{delplace2017topological,graf2021topology}.

Finally, we investigate the robustness of the spectral flow in the presence of
non-Hermitian perturbations arising from collisional damping.
We show that the spectral flow of the real parts of the eigenvalues remains
well defined and quantized as long as a finite strip gap is preserved and no
exceptional points enter this gap~\cite{gong2018topological,kawabata2019symmetry}.
This analysis delineates the regime of applicability of the bulk--interface
correspondence in realistic dissipative plasma systems and highlights the
topological origin of chiral wave transport beyond idealized Hermitian models.

\section{Wave Topology of Screened Magnetized Plasma}
In this section, we develop the bulk topological characterization of a
screened magnetized plasma at the level of its phase-space symbol.
Our goal is to identify the relevant band degeneracies of the Weyl symbol,
define the associated strip-gap Chern numbers, and establish the bulk
topological invariants that underlie the bulk--interface correspondence
discussed in the Introduction.
We first cast the linearized plasma dynamics into a pseudo-Hermitian
Schrödinger-type form, and then analyze the topology of the resulting Weyl
symbol in phase space.

With a uniform background magnetic field $\boldsymbol{B}_0$ and vanishing background flow, 
the linearized dynamics of a magnetized cold plasma is governed by
\begin{eqnarray}\label{eqn1}
\begin{gathered}
    m_e\frac{\partial\boldsymbol{v}_1}{\partial t}
    =e\nabla\phi_1+e\boldsymbol{v}_1\times \boldsymbol{B}_0,\\
    \frac{\partial n_1}{\partial t}
    =-n_0\nabla\cdot\boldsymbol{v}_1,\\
    (\nabla^2-\lambda^2)\phi_1
    =4\pi en_1 .
\end{gathered}
\end{eqnarray}
Here the displacement current is neglected, and the electric field is assumed to arise solely from the perturbed electron density $n_1$.
$\boldsymbol{v}_1$ denotes the perturbed electron velocity, $\lambda^{-1}$ is the screening length of the plasma,
and $\phi_1$ is the electrostatic potential.
The electron charge and mass are denoted by $e$ and $m_e$, respectively, and $n_0$ is the uniform background density.

We normalize time by the inverse plasma frequency $\omega_{pe}^{-1}$,
with $\omega_{pe}=\sqrt{4\pi n_0 e^2/m_e}$,
and length by the screening length $\lambda^{-1}$.
The electrostatic potential is normalized by $4\pi n_0 |e|/\lambda^2$,
and the velocity by $\omega_{pe}/\lambda$.
Introducing the state vector
\[
\Xi=(\boldsymbol{v}_1,\phi_1)^{\mathsf T},
\]
and defining the dimensionless cyclotron frequency
$\boldsymbol{\mu}=(|e|\boldsymbol{B}_0/m_e)/\omega_{pe}$,
the system \eqref{eqn1} can be cast into a generalized Schrödinger-type form~\cite{mostafazadeh2002pseudo,jin2016topological}
\begin{equation}\label{pdesch}
    \mathrm{i}\hat{\eta}\,\partial_t\Xi=\hat{H}\,\Xi.
\end{equation}
Here $\hat{H}$ is a Hermitian operator, while $\hat{\eta}$ is a positive-definite Hermitian metric operator,
\begin{equation}
    \hat{H}=
    \begin{pmatrix}
    \mathrm{i}\boldsymbol{\mu}\times & -\mathrm{i}\nabla\\
    -\mathrm{i}\nabla\cdot & 0
    \end{pmatrix},
    \qquad
    \hat{\eta}=
    \begin{pmatrix}
    I & 0\\
    0 & -\nabla^2+1
    \end{pmatrix}.
\end{equation}
Equation \eqref{pdesch} therefore defines a pseudo-Hermitian dynamical system,
in which the generator $\hat{\eta}^{-1}\hat{H}$ is similar to a Hermitian operator
and possesses a purely real spectrum as long as $\hat{\eta}$ remains positive definite~\cite{mostafazadeh2002pseudo}.

We consider inhomogeneous magnetic configurations of the form
$\boldsymbol{\mu}=\mu(x)\boldsymbol{e}_z$,
varying only along the $x$ direction.
Under this assumption, a Fourier transform can be performed along the homogeneous directions $(y,z,t)$,
such that the transverse wave numbers $k_y$ and $k_z$ are conserved parameters.
The resulting eigenvalue problem along $x$ thus depends parametrically on $(k_y,k_z)$.

For continuous media, the spectral and topological properties are encoded in the phase-space structure of the associated Weyl symbol~\cite{wigner1932quantum,faure2019manifestation}.
Applying the Wigner transform along the $x$ direction,
we obtain the exact Weyl symbol of the effective generator $\hat{\eta}^{-1}\hat{H}$,
\begin{equation}\label{symbol}
\mathrm{sym}(\hat{\eta}^{-1}\hat{H})=
\begin{pmatrix}
0 & -\mathrm{i}\mu(x) & 0 & k_x\\
\mathrm{i}\mu(x) & 0 & 0 & k_y\\
0 & 0 & 0 & k_z\\
\dfrac{k_x}{k^2+1} & \dfrac{k_y}{k^2+1} & \dfrac{k_z}{k^2+1} & 0
\end{pmatrix},
\end{equation}
where $k^2=k_x^2+k_y^2+k_z^2$.
The symbol satisfies the pseudo-Hermitian symmetry relation
\[
\mathrm{sym}(\hat{\eta})\,\mathrm{sym}(\hat{\eta}^{-1}\hat{H})\,\mathrm{sym}(\hat{\eta}^{-1})
=\mathrm{sym}(\hat{\eta}^{-1}\hat{H})^{\dagger},
\]
and the positivity of the screening-induced metric $\hat{\eta}$ guarantees that this symmetry is unbroken. Let the symbol of \( \hat{\eta} \) be denoted by \( \eta \), and the symbol of \( \hat{H} \) by \( H \). Then, it is evident that
\begin{equation}\label{symvanish}
    \mathrm{sym}(\hat{\eta}^{-1}\hat{H}) = \eta^{-1} H.
\end{equation}
Strictly speaking,
\begin{equation}
\mathrm{sym}(\hat{\eta}^{-1} \hat{H}) = \mathrm{sym}(\hat{\eta}^{-1}) \star H,
\end{equation}
where \( \star \) denotes the Moyal product. However, \( \hat{\eta}^{-1} \) acts only on the fourth row of the operator \( \hat{H} \), and it is independent of the parameter \( x \). The \( x \)-dependent terms appear only in \( \mu \), which enters the first and second rows of \( \hat{H} \). Therefore, the equality
\begin{equation}
\mathrm{sym}(\hat{\eta}^{-1} \hat{H}) = \eta^{-1} H \label{symvanish}
\end{equation}
holds exactly.
\begin{figure}
    \centering
    \includegraphics[width=0.5\textwidth]{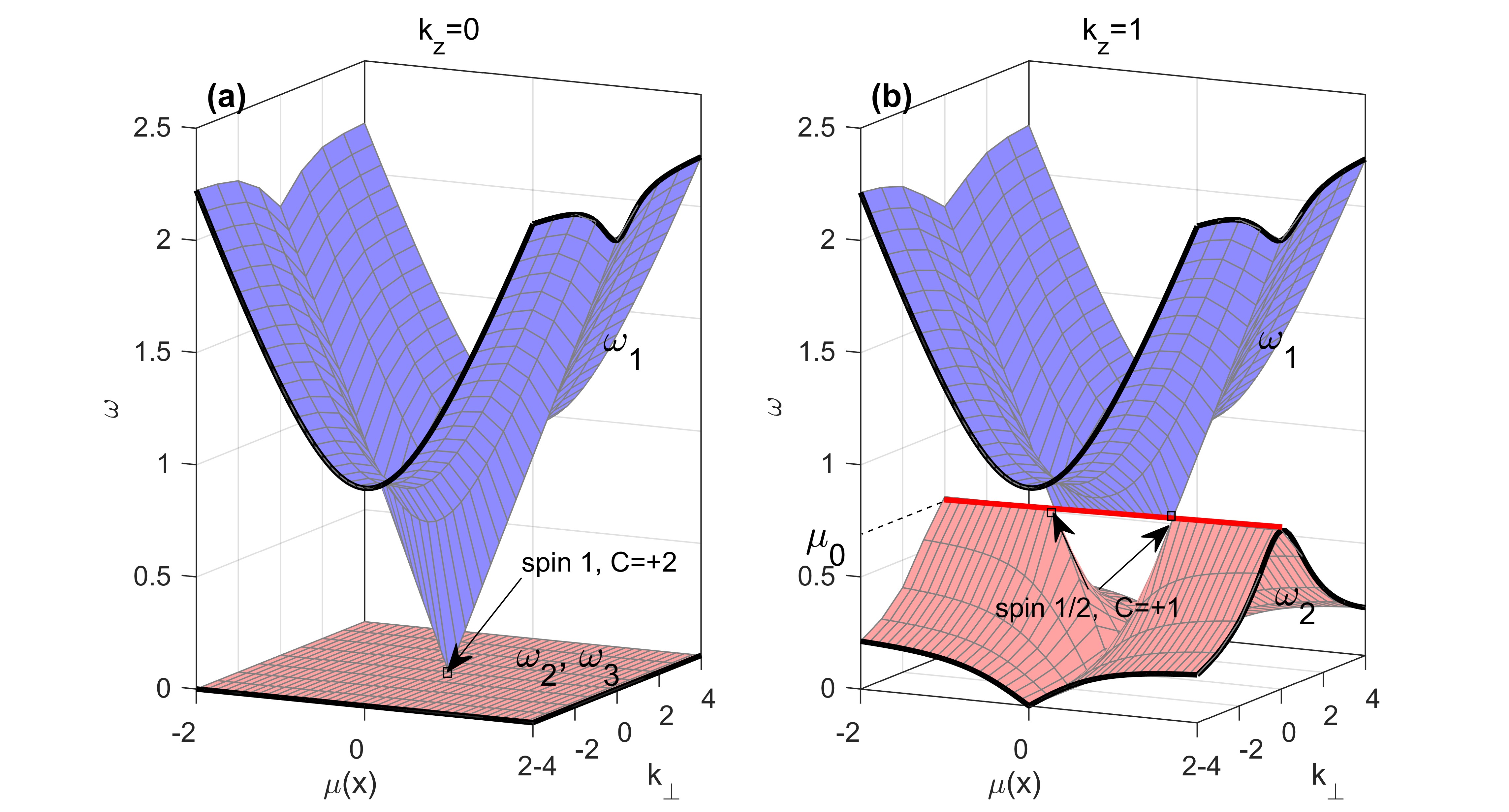}
   \caption{
Schematic illustration of the bulk band structure of the Weyl symbol
$\eta^{-1}H$ in the parameter space $(\mu,k_x,k_y)$ for fixed $k_z$.
Isolated band degeneracies appear as Weyl points (marked by arrows),
each carrying a topological charge.
(a) For $k_z=0$, the spectrum exhibits a single higher-order Weyl point
with topological charge $+2$.
(b) For $k_z\neq0$, this degeneracy splits into two Weyl points, each
carrying a unit topological charge $+1$.}

    \label{fig1}
\end{figure} 

Before proceeding to the interface problem, we summarize the role of the bulk
analysis presented above.
Crucially, the bulk topology is entirely encoded in the Weyl symbol
$\eta^{-1}H(\mu,k_x,k_y,k_z)$ and its isolated band degeneracies in the extended
parameter space $(\mu,k_x,k_y)$.
The associated gap Chern numbers provide the topological input for the
bulk--interface correspondence developed in the next section.

The Weyl symbol $\eta^{-1}H(\mu,k_x,k_y,k_z)$ exhibits isolated band
degeneracies (Weyl points) in the extended parameter space
$(\mu,k_x,k_y)$, whose topological charges are characterized by the
gap Chern number~\cite{delplace2017topological,jezequel2023mode}
defined in Eq.~(\ref{c-gap}).
In the following section, we show that these bulk topological invariants
govern the spectral flow of interface modes when the reduced magnetic
field varies spatially between two asymptotic values.

To this end, we allow the reduced magnetic field to depend on the
spatial coordinate, $\mu=\mu(x)$, and assume that $\mu(x)$ is smooth
and approaches constant limits as $x\to\pm\infty$.
Under this assumption, the topology of the symbol in phase space
$(x,k_x,k_y)$ is determined by the corresponding bulk symbols evaluated
at the asymptotic values of $\mu$.

Throughout this work, a \emph{strip gap} refers to a real-frequency
interval $I_\omega\subset\mathbb{R}$ such that, for all sufficiently
large $\|(x,k_x,k_y)\|$, the spectrum of the Weyl symbol
$\eta^{-1}H(x,k_x,k_y)$ does not intersect $I_\omega$,
except possibly at isolated band-degeneracy points.

This notion generalizes the concept of a band gap to continuous media
with unbounded phase space, where the spectrum typically extends to
infinity and compact Brillouin-zone arguments no longer apply
~\cite{silveirinha2015chern,parker2020topological}.

The symbol $\mathrm{sym}(\hat{\eta}^{-1}\hat{H})$ possesses four real eigenvalues,
which can be written compactly as
\begin{widetext}
\[
(\pm)_1
\frac{
\sqrt{
k^{2}+\left(1+k^{2}\right)\mu^{2}
+(\pm)_2\sqrt{\left(k^{2}+\left(1+k^{2}\right)\mu^{2}\right)^{2}
-4\left(1+k^{2}\right)k_{z}^{2}\mu^{2}}
}
}{\sqrt{2}\sqrt{1+k^{2}}}.
\]
\end{widetext}
The four branches are ordered as $\omega_1\ge\omega_2\ge\omega_3\ge\omega_4$,
with $\omega_1=-\omega_4$ and $\omega_2=-\omega_3$.
For fixed $k_z$ and $\mu$, the asymptotic behavior at large $k$ is
\[
\{\omega_1,\omega_2,\omega_3,\omega_4\}
\to
\{\sqrt{1+\mu^2},\,0,\,0,\,-\sqrt{1+\mu^2}\}.
\]

When $k_z=0$, the two middle bands $\omega_2$ and $\omega_3$ become completely degenerate at zero frequency.
In this limit, the symbol contains an exact zero row and zero column,
reducing the problem effectively to a three-band system.
All four bands meet at a single degeneracy point located at
$(\mu,k_x,k_y)=(0,0,0)$, as illustrated in Fig.~\ref{fig1}(a).
This fourfold degeneracy represents a higher-spin (spin-1) band touching,
which goes beyond the conventional spin-$\tfrac{1}{2}$ Weyl point and
provides a natural extension of bulk--interface correspondence to
higher-spin pseudo-Hermitian systems.

When $k_z\neq0$, this higher-order degeneracy is lifted.
The original spin-1 degeneracy splits into two isolated spin-$\tfrac{1}{2}$
Weyl points located at
\[
(\mu,k_x,k_y)=(\pm\mu_0,0,0),
\qquad
\mu_0=\sqrt{\frac{k_z^2}{1+k_z^2}} ,
\]
as illustrated in Fig.~\ref{fig1}(b).
Physically, a finite $k_z$ plays the role of a symmetry-breaking parameter that
lifts the spin-1 degeneracy and splits it into two spin-$\tfrac{1}{2}$ Weyl
points, in close analogy with the splitting of a double Weyl point in
topological semimetals under symmetry reduction.

To characterize the topological properties of the spectrum away from isolated
band degeneracies, we introduce the Chern number associated with a spectral
gap.
For a given frequency gap separating the $n$-th and $(n+1)$-th bands, the gap
Chern number is defined as
\begin{equation}\label{c-gap}
C_{\mathrm{gap}}
=
\frac{i}{2\pi}
\int_{\Sigma}
\mathrm{Tr}
\left(
P\,\mathrm{d}P\wedge \mathrm{d}P
\right),
\end{equation}
where $P$ denotes the projector onto all bands below the gap, and $\Sigma$ is a closed two-dimensional surface in the extended parameter space $(\mu,k_x,k_y)$ that does
not intersect any band degeneracy. In the integral on the right-hand side of Eq.~(\ref{c-gap}), when  $\Sigma$ encloses only a single Weyl point, the result corresponds to the topological charge of that individual Weyl point. When $\Sigma$ encloses all the Weyl points within a given gap, the integral yields the gap Chern number. Owing to the closedness of the integrand, the Stokes theorem ensures that the gap Chern number equals the sum of the topological charges of all Weyl points within the gap.

We now proceed to evaluate Eq.~(\ref{c-gap}) to verify the previous assertion regarding the specific values of the topological charges. Let the right and left eigenvectors of \( \mathrm{sym}(\hat{\eta}^{-1}\hat{H}) \) corresponding to the eigenvalue \( \omega_i \) be denoted as \( |\Psi^R_i\rangle \) and \( \langle\Psi^L_i| \), respectively. Then,
\begin{align}
\eta^{-1}H\, |\Psi^R_i\rangle 
&= \omega_i\, |\Psi^R_i\rangle, \\
\langle\Psi^L_i|\, \eta^{-1}H
&= \omega_i\, \langle\Psi^L_i|.
\end{align}
Due to the presence of pseudo-Hermitian symmetry with respect to \( \hat{\eta} \), the eigenvectors can be chosen to satisfy the biorthogonality condition~\cite{mostafazadeh2002pseudo}:
\begin{equation}
\langle \Psi^L_i | \Psi^R_j \rangle = \delta_{ij},
\end{equation}
and
\begin{equation}
| \Psi^L_i \rangle =  \eta| \Psi^R_i \rangle.
\end{equation}
Accordingly, the projector \( P \) onto the gap between \( \omega_1 \) and \( \omega_2 \) can be expressed as:
\begin{equation}
P = \sum_{i=2}^{4} |\Psi^R_i\rangle \langle \Psi^L_i|.
\end{equation}

Using the eigenvalue equations satisfied by \( |\Psi^{R/L}_i\rangle \), namely
\begin{align}
\omega_i |\Psi^R_i\rangle &= \eta^{-1} H |\Psi^R_i\rangle, \\
\omega_i |\Psi^L_i\rangle &= (\eta^{-1} H)^\dagger |\Psi^L_i\rangle,
\end{align}
one can derive the expression for the gap Chern number:
\begin{eqnarray*}
C_{\mathrm{gap}} &= -\frac{i}{2\pi}\sum_{i=2}^{4} \int_{\Sigma} \mathrm{Im} \left(  \sum_{j \neq i} \frac{ \langle \Psi^L_i | \nabla H | \Psi^R_j \rangle \times \langle \Psi^L_j | \nabla H | \Psi^R_i \rangle }{(\omega_i - \omega_j)^2} \right)\\
&=:\sum_{i=2}^{4} C_i
\end{eqnarray*}
where the surface integral is taken over a closed surface \( \Sigma \) in phase space $(x,k_x,k_y)$ enclosing degenaracies of band 1 and band 2. We denote \(C_i\) the Chern number of band i corresponding to eigenvalue \(\omega_i\). Similarly, the gap  Chern number for gap between band \(j-1\) and band \(j\) is \(\sum_{i=j}^{4} C_i\). The expressions of \(C_i\)s correspond to the Berry-curvature integral for bands of a pseudo-Hermitian system. One can see that, when \( \eta \) continuously approaches the identity matrix \( I \), \(\sum_{i=j}^{4} C_i\) smoothly reduces to the gap Chern number in the Hermitian case. Therefore, the topology associated with a pseudo-Hermitian symbol is homotopic to that of its Hermitian counterpart. Because the metric operator $\eta$ is positive definite and varies smoothly in
the phase space $(x,k_x,k_y)$, the associated pseudo-Hermitian vector
bundle is homotopic to a Hermitian vector bundle.
As a result, the Berry curvature and Chern numbers can be computed using the
biorthogonal eigenvectors without introducing additional metric-connection
corrections.

Numerical integration shows that, for \( k_z = 0 \), the degenerate point contributes a topological charge of \( +2 \). When \( k_z \neq 0 \), the two degenerate points each contribute a topological charge of \( +1 \). In other words, the total topological charge between band~1 and band~2 is conserved. The above conservation law of the gap Chern number, which is independent of \( k_z \), holds for any gap between band \( j-1 \) and band \( j \).

Near $k=0$, where $\hat{\eta}\simeq I$, the local dynamics reduces to that of the
shallow-water equations describing equatorial geostrophic waves, and the
degeneracy carries a topological charge $C=+2$, originating from two
co-propagating chiral modes\cite{delplace2017topological}.
\section{Bulk interface correspondence for screened magetized plasma}
The gap Chern number introduced in Eq.~\ref{c-gap} determines the net spectral flow of the interface eigenvalues as a function of $k_y$. In particular, as $k_y$ varies from $-\infty$ to $+\infty$, it counts the net number of discrete eigenvalue branches that detach from the continuous spectrum associated with $\omega_2$, traverse the strip gap, and reconnect with the continuous spectrum associated with $\omega_1$.

We now consider a configuration in which the reduced magnetic field $\mu(x)$ varies along the spatial $x$-direction, while the transverse wave numbers $k_y$ and $k_z$ are held fixed. We assume that the profile $\mu(x)$ is such that the symbol $\eta^{-1} H$ admits a common strip gap for sufficiently large $\|(x, k_x, k_y)\|$. Under this condition, the problem reduces to a one-dimensional eigenvalue problem parameterized by $k_y$, whose spectral flow we analyze.

Importantly, the behavior differs qualitatively between the cases $k_z = 0$ and $k_z \neq 0$, as well as for different asymptotic values $\mu(\pm\infty)$. This distinction stems from the structure of the bulk degeneracies: when $k_z \neq 0$, the single degenerate point carrying a topological charge of $+2$ splits into two isolated degeneracies, each carrying unit charge $+1$.

To demonstrate this behavior, we solve the associated eigenvalue problem
on a finite interval $[-L,L]$ with Dirichlet boundary conditions.
We adopt the smooth profile
\begin{equation}\label{mux}
\mu(x) = \frac{\mu_1 + \mu_2}{2}
+ \frac{\mu_2 - \mu_1}{2}
\tanh\!\left(\frac{x}{\delta}\right),
\end{equation}
which interpolates between the asymptotic values
$\mu_1$ and $\mu_2$.]

For \( k_z = 0 \), the parameters \( \mu_1 \) and \( \mu_2 \) can be chosen as arbitrary nonzero values such that the common strip gap condition discussed above is satisfied.
By contrast, when $k_z \neq 0$,
a common strip gap exists only if
\begin{equation}
\mu_1, \mu_2 \in (-\mu_0, +\mu_0)
\quad \text{or} \quad
\mu_1, \mu_2 \notin [-\mu_0, +\mu_0].
\end{equation}
This restriction can be understood
from the bulk structure shown in Fig.~\ref{fig1}.

When the bulk--interface correspondence, equivalently the spectral-flow--monopole correspondence, holds in the present model, the following behavior is expected.

For \( k_z = 0 \), the spectral flow within the strip gap equals 2~\(\text{sign}(\mu_2-\mu_1)\) when
\begin{equation}
\mu_1 \mu_2 < 0,
\end{equation}
and vanishes when
\begin{equation}
\mu_1 \mu_2 > 0.
\end{equation}

For \( k_z \neq 0 \), the spectral flow equals 2~\(\text{sign}(\mu_2-\mu_1)\) when one of \( \mu_{1,2} \) satisfies
\begin{equation}
\mu_{1/2} < -\mu_0,
\end{equation}
while the other satisfies
\begin{equation}
\mu_{2/1} > \mu_0.
\end{equation}
When
\begin{equation}
-\mu_0 < \mu_{1/2} < \mu_0,
\end{equation}
the spectral flow vanishes.

In contrast, when one of \( \mu_{1,2} \) lies in the interval \( (-\mu_0, \mu_0) \) while the other lies outside
\( [-\mu_0, \mu_0] \), the strip gap condition is violated at
\( (k_x, k_y) = (0,0) \) for arbitrary values of \( \mu(x) \).
Although the system crosses a spin-$\tfrac{1}{2}$ degeneracy carrying unit topological charge under the continuous deformation of $\mu(x)$ from $\mu_1$ to $\mu_2$, no well defined spectral flow can be assigned within the strip gap.
In this situation, the strip gap is violated only at $(k_x,k_y) = (0,0)$ for sufficiently large $\|(x,k_x,k_y)\|$. Consequently, for $k_y \neq 0$ the strip gap remains open, and nontrivial topology may still persist.

\subsection{Numerical results and spectral flow}
We now present the numerical results for the interface eigenvalue problem and demonstrate the spectral flow predicted by the gap Chern number.

For fixed transverse wave numbers \( k_y \) and \( k_z \), we solve the generalized eigenvalue problem
\begin{equation}
\hat{H}(k_y,k_z;\mu(x))\,\Psi(x)
= \omega\,\hat{\eta}(k_y,k_z)\,\Psi(x),
\end{equation}
on a finite interval \( x\in[-L,L] \), subject to Dirichlet boundary conditions. The Dirichlet boundary condition is imposed solely for numerical convenience. We have verified that the existence and chirality of the interface modes are insensitive to the specific choice of boundary conditions, provided that the system size $L$ is sufficiently large.
Here \(\hat{H}(k_y,k_z;\mu(x))\) denotes the one-dimensional differential operator obtained by replacing
\( k_x \rightarrow -\mathrm{i}\partial_x \) in the bulk symbol \( H \), and
\(\hat{\eta}\) is the corresponding metric operator.
The spatial profile \( \mu(x) \) is chosen as in Eq.~(\ref{mux}), varying smoothly between the
asymptotic values \( \mu_1 \) and \( \mu_2 \).
The system size \( L \) is taken to be much larger than the interface width \( \delta \),
so that the boundary does not affect the localized interface modes.
We have verified that the spectral features reported below are insensitive to further increases of \( L \).

\begin{figure}
    \centering
    \includegraphics[width=1\linewidth]{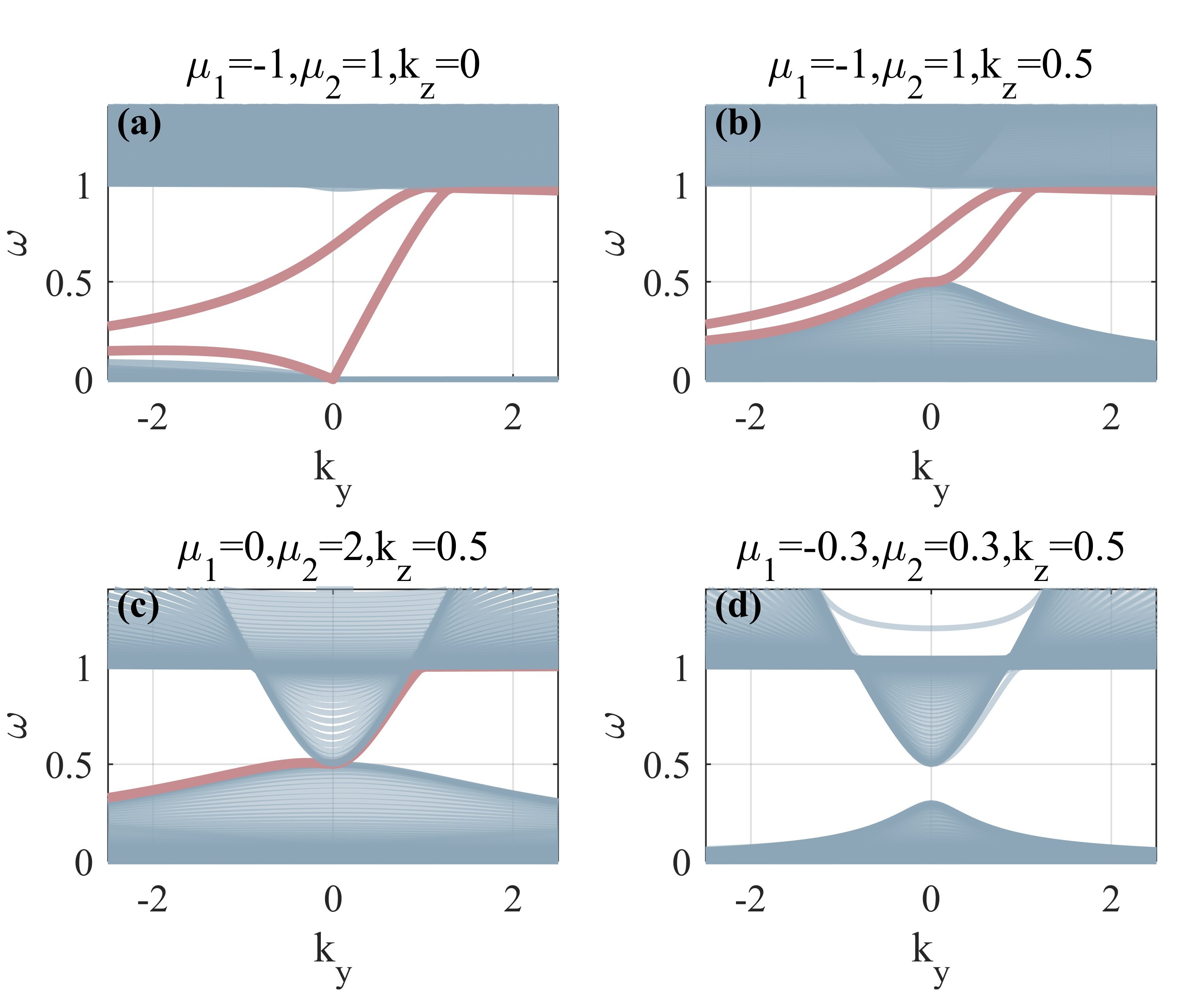}
    \caption{
Eigenvalue spectra of the interface problem as functions of $k_y$ for
different magnetic configurations $\mu(x)$.
Discrete eigenvalues inside the strip gap are marked by plus symbols.
(a,b) Cases with a common strip gap.
For $k_z=0$, the variation of $\mu(x)$ crosses a single higher-order
degeneracy carrying topological charge $+2$; for $k_z\neq0$, it crosses two
Weyl points each carrying unit topological charge $+1$.
In both cases, two chiral branches traverse the strip gap, yielding a net
spectral flow of $+2$, consistent with the gap Chern number.
(c) Case without a strip gap at $k_y=0$.
Although the gap is locally closed, the variation crosses a single Weyl
point with unit topological charge, resulting in a single chiral branch
connecting the two continuous spectra.
(d) Case with a strip gap but no band degeneracy crossed during the
variation, for which no discrete interface modes appear within the gap.
}

    \label{fig2}
\end{figure}

Figure~\ref{fig2} shows the eigenvalue spectrum as a function of \( k_y \)
for a representative interface configuration.
Within the strip gap, two discrete branches emerge from the bulk continuum,
traverse the gap monotonically, and merge into the opposite bulk band as \( k_y \) varies
from \( -\infty \) to \( +\infty \).
For cases in which a strip gap exists, when \( k_z = 0 \) and \( \mu(x) \) varies from \( \mu_1 \) to \( \mu_2 \),
a degenerate point of symbol \(\eta^{-1}H\) with topological charge \( +2 \) appears near the interface.
Alternatively, when \( k_z \neq 0 \), the interface contains two degenerate points,
each carrying a topological charge \( +1 \).
In both cases, a spectral flow of \( +2 \) can be observed within the strip gap,
as shown in Fig.~2(a,b).
The plus symbols indicate discrete eigenvalues in the gap flowing from the lower continuous spectrum to the upper continuous spectrum.
The net number of gap-crossing modes is equal to two, in agreement with the gap Chern number predicted from the bulk--interface correspondence.

Figure~2(c) illustrates a relatively ``anomalous'' phenomenon.
In this case, the system does not possess a strip gap at \( k_y = 0 \);
nevertheless, the numerical simulation still reveals a chiral branch
that connects the two continuous spectra.
Notably, as \( \mu(x) \) varies from \( \mu_1 \) to \( \mu_2 \),
the system crosses a single degenerate point carrying a topological charge of \( +1 \).
Therefore, this chiral spectrum can also be regarded as consistent with the bulk--interface correspondence. 
From a topological perspective, this behavior does not contradict the
bulk--interface correspondence.
Although the strip gap is locally closed at $(k_x,k_y)=(0,0)$, the variation
of $\mu(x)$ still crosses a single Weyl point carrying unit topological charge,
which enforces a chiral spectral flow.
This illustrates the robustness of monopole-induced spectral flow against local
violations of the strip-gap condition.

Figure~2(d) shows that, when the Weyl symbol does not exhibit any degeneracy
at the interface and the system possesses a strip gap,
the corresponding eigenvalue spectrum contains no discrete modes within the gap.

\section{Spectral flow under collisional damping}

In realistic plasma systems, collisional effects and dissipation inevitably
introduce damping into the electron dynamics.
Such effects render the governing equations non-Hermitian and break the
pseudo-Hermitian symmetry underlying the idealized model discussed above.
It is therefore natural to ask whether the spectral flow predicted by the gap
Chern number remains robust in the presence of damping.
In this section, we address this question by incorporating a simple collisional
damping term and analyzing its impact on the interface spectrum.

\subsection{Damped model and non-Hermitian perturbation}

We introduce a phenomenological damping rate $\nu>0$ into the electron momentum
equation,
\begin{equation}
m_e\frac{\partial\boldsymbol{v}_1}{\partial t}
=
e\nabla\phi_1
+ e\boldsymbol{v}_1\times\boldsymbol{B}_0
- m_e\nu\,\boldsymbol{v}_1 ,
\end{equation}
while keeping the continuity equation and the screened Poisson equation
unchanged.
After normalization using the same conventions as in Sec.~II, the resulting
generalized Schrödinger-type equation takes the form
\begin{equation}
\mathrm{i}\hat{\eta}\,\partial_t\Xi
=
\left(
\hat{H}
+ \mathrm{i}\nu \hat{D}
\right)\Xi ,
\end{equation}
where $\hat{H}$ and $\hat{\eta}$ are the Hermitian operators defined previously,
and
\begin{equation}
\hat{D}
=
\begin{pmatrix}
- I & 0\\
0 & 0
\end{pmatrix}
\end{equation}
acts only on the velocity components.
The additional term $\mathrm{i}\nu\hat{D}$ explicitly breaks the
pseudo-Hermitian symmetry and renders the spectrum generally complex.

\subsection{Persistence and breakdown of spectral flow}

In the absence of damping ($\nu=0$), the interface spectral flow is protected by
the gap Chern number defined in the phase space $(x,k_x,k_y)$,
and the system is pseudo-Hermitian with a purely real spectrum.
When $\nu$ is finite, the eigenvalues evolve continuously from their values at $\nu=0$, provided that no exceptional points are encountered, in accordance with the general theory of non-Hermitian spectral topology~\cite{gong2018topological,kawabata2019symmetry}.
In this regime, each real eigenvalue $\omega$ acquires a finite imaginary part,
while its real part varies continuously with respect to both $\nu$ and $k_y$.

\begin{figure*}
    \centering
    \includegraphics[width=1\textwidth]{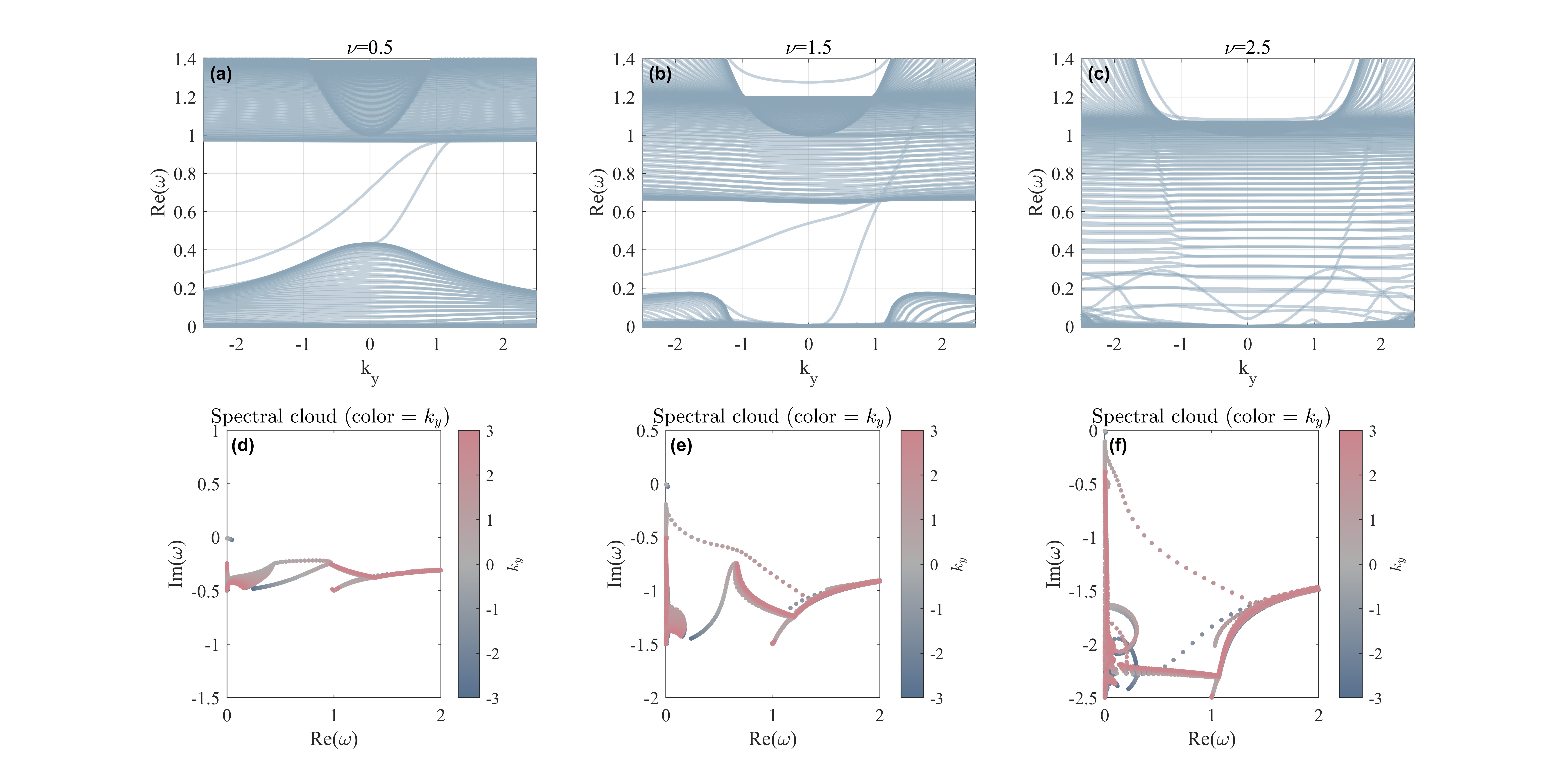}
    \caption{
Interface eigenvalue spectra in the presence of finite damping for
$\nu=0.5$, $1.5$, and $2.5$.
Panels (a--c) show the real parts of the eigenfrequencies as functions of
$k_y$, while panels (d--f) display the corresponding spectral clouds in the
complex-frequency plane.
(a,d) For $\nu=0.5$, complex eigenfrequencies appear, but the real parts of the
spectrum retain a finite strip gap, within which two discrete branches
traverse between the two continuous spectra.
(b,e) For $\nu=1.5$, the strip gap in the real spectrum is reduced, and the
separation between the two continuous spectra in the spectral cloud becomes
narrower, while the gap-crossing branches remain identifiable.
(c,f) For $\nu=2.5$, the strip gap closes completely and the two continuous
spectra merge in the real-frequency projection; the spectral cloud consists
of two families of discrete eigenvalues parameterized by $k_y$, corresponding
to the continuation of the gap-crossing branches observed at smaller damping.
}

    \label{fig3}
\end{figure*}

As long as the real parts of the spectrum continue to satisfy the strip-gap condition, the spectral flow of
the real parts of the eigenvalues remains well defined.
In particular, the net number of eigenvalue branches whose real parts traverse
the strip gap as $k_y$ varies from $-\infty$ to $+\infty$ is unchanged from the
Hermitian case and continues to coincide with the gap Chern number.

When the damping becomes sufficiently strong, however, exceptional points may
emerge and enter the strip gap.
In this situation, eigenvalues can no longer be continuously tracked, and the
notion of spectral flow within the strip gap breaks down.
This signals the loss of topological protection and marks the boundary of
applicability of the bulk--interface correspondence in the presence of strong
non-Hermitian effects.

\subsection{Numerical illustration}

To substantiate the above analysis, we perform numerical calculations for
finite damping rates $\nu=0.5,\,1.5,$ and $2.5$.
Throughout this subsection, we examine how weak to moderate damping affects
the interface spectrum and the associated spectral flow.

For $\nu=0.5$, complex eigenfrequencies already appear, as shown in
Fig.~\ref{fig3}(a).
Nevertheless, the real parts of the spectrum still exhibit a well-defined
strip gap.
Correspondingly, in the complex-frequency plane [Fig.~\ref{fig3}(d)], the spectral
cloud clearly reveals two discrete branches traversing between the two
continuous spectra as $k_y$ varies.
These branches represent a spectral flow that remains identical to the
Hermitian case, despite the presence of finite imaginary parts.

When the damping is increased to $\nu=1.5$, the strip gap in the real part of
the spectrum becomes noticeably narrower, as shown in Fig.~\ref{fig3}(b).
The corresponding spectral cloud in Fig.~\ref{fig3}(e) indicates that the separation
between the two continuous spectra is reduced.
Nevertheless, two gap-crossing branches can still be identified, indicating
that the spectral flow persists as long as a finite strip gap remains in the
real spectrum.

For a larger damping rate $\nu=2.5$, the strip gap closes completely.
As shown in Fig.~\ref{fig3}(c), the two continuous spectra merge in the real-frequency
projection.
In this regime, the spectral cloud [Fig.~\ref{fig3}(f)] consists of two families of
discrete eigenvalues parameterized by $k_y$, which can be viewed as the
continuation of the gap-crossing branches observed at smaller damping.
However, since no strip gap exists in the real spectrum, a well-defined
spectral flow within a gap can no longer be assigned.

These numerical results illustrate that the spectral flow predicted by the
gap Chern number remains robust under finite damping, as long as the real parts
of the spectrum retain a strip gap.
Once the gap is destroyed, the notion of spectral flow within the strip gap
breaks down, signaling the loss of topological protection.

\section{Conclusion}

We have established a unified phase-space topological framework for wave dynamics in screened magnetized plasmas based on a pseudo-Hermitian formulation with a positive-definite metric. By recasting the linearized plasma equations into a generalized Schrödinger form, we identified a mathematically controlled setting in which bulk topology can be defined despite the absence of a compact Brillouin zone and conventional band gaps.

At the level of the Weyl–Wigner symbol, we showed that the bulk spectrum hosts isolated band degeneracies in phase space that act as Berry–Chern monopoles carrying quantized topological charges. In particular, a higher-order spin-1 degeneracy with charge $+2$ generically splits into two spin-$\tfrac{1}{2}$ Weyl points under symmetry breaking, while preserving the total strip-gap topological charge. These monopoles provide the microscopic origin of the spectral flow observed in the interface problem.

To characterize topology away from such degeneracies in a continuous medium with unbounded phase space, we introduced the strip-gap Chern number. This invariant extends the notion of band Chern number to continuous media by assigning topology to finite real-frequency strips rather than isolated bands. It governs the spectral flow of eigenvalues across the strip gap and provides a robust topological characterization of continuum wave systems.

By solving the interface eigenvalue problem generated by a smooth spatial variation of the magnetic field, we demonstrated that the net number of interface eigenvalue branches whose real parts traverse the strip gap is fully determined by the total enclosed topological charge. Our numerical results confirm this bulk–interface correspondence across a broad parameter range, including situations in which the strip gap is locally violated at isolated points in phase space. This clarifies both the monopole-induced origin of spectral flow and the role of higher-order degeneracies in continuous plasma systems.

We further examined the robustness of this correspondence in the presence of non-Hermitian effects arising from collisional damping. Although dissipation renders the spectrum complex and breaks pseudo-Hermitian symmetry, the spectral flow of the real parts of the eigenvalues remains quantized as long as a finite strip gap persists and no exceptional points enter this gap. Once the strip gap closes, eigenvalues can no longer be continuously tracked, and the notion of spectral flow within the gap ceases to be well defined. This sharply delineates the regime in which a strip-gap-based topological description remains meaningful in dissipative continua.

More broadly, while topological wave phenomena in continuous media have been explored in various contexts, our results provide a concrete phase-space framework in which pseudo-Hermiticity, strip-gap topology, and bulk–interface correspondence are treated on equal footing. The formulation developed here can in principle be extended to fluid, plasma, and electromagnetic systems with constraints and long-range interactions, and may serve as a foundation for understanding robust chiral wave transport beyond idealized lattice models.

\env{acknowledgments} 
 This work was supported by the National Natural Science Foundation of China [Grant No. 11775220].

\bibliography{reference}

\end{document}